\documentclass[aps,showpacs,manuscript,12pt]{revtex4}
\usepackage{amssymb}
\usepackage{amsmath}
\usepackage{graphicx}

\begin{document}

\title{\textbf{Lagrangian dynamics of incompressible thermofluids$^{\S }$}}
\author{Marco Tessarotto$^{a,b,c}$, Claudio Cremaschini$^{d}$, Piero Nicolini$^{c,e}$
and Massimo Tessarotto$^{c,e}$} \affiliation{\ $^{a}$Civil
Protection Agency, Regione Friuli Venezia-Giulia, Palmanova,
Italy, $^{b}$Department of Electronics, Electrotechnics and
Informatics, University of Trieste, Italy, $^{c}$Consortium of
Magneto-fluid-dynamics, University of Trieste, Italy,
$^{d}$Department of Astronomy, University of Trieste, Italy,
$^{e}$Department of Mathematics and Informatics, University of
Trieste, Italy}

\begin{abstract}
A key aspect of fluid dynamics is the correct definition of the \textit{%
phase-space} Lagrangian dynamics which characterizes arbitrary
fluid elements of an incompressible fluid. Apart being an unsolved
theoretical problem of fundamental importance, the issue is
relevant to exhibit the connection between fluid dynamics and the
classical dynamical systems underlying incompressible and
non-isothermal fluid, typically founded either on: a) a
\textit{configuration-space} Lagrangian description of the
dynamics of fluid elements; b) a kinetic description of the
molecular dynamics, based on a discrete representation of the
fluid. The goal of this paper is to show that the exact Lagrangian
dynamics can be established based on the inverse kinetic theory
(IKT) for incompressible fluids recently pointed out (Ellero
\textit{et al.}, 2004-2006, \cite{Ellero2004}). The result is
reached by adopting an IKT approach based on a \textit{restricted
phase-space representation} of the fluid, in which the
configuration space coincides with the physical fluid domain. The
result appears of potential importance in applied fluid dynamics
and CFD.
\end{abstract}

\date{\today }
\maketitle



\section{Introduction}

As it is well known, the description of fluids can be performed
choosing either an Eulerian or a Lagrangian point of view. The two
approaches are equivalent and, if fluid dynamics were fully
understood, one should be able to translate Eulerian properties
into Lagrangian ones and viceversa. At least for the treatment of
turbulent fluids, we are still quite far from this point
\cite{Richardson}. The main historical reason of this situation
can be understood by looking at the customary statistical approach
based on so-called velocity probability density function (pdf)-
method for an incompressible fluid (for a review see for example
\cite{Pope2000}). In fact - as it is well known - in this approach
the time evolution of the pdf which advances in time the average
(in a stochastic sense) fluid velocity is determined by a
Fokker-Planck transport equation. As a consequence the
corresponding Lagrangian characteristics are necessarily
stochastic in nature and therefore difficult to handle. This
explains why in the literature, for this purpose, stochastic
models of various nature have been adopted, which, however,
typically rely on experimental verification rather than (uniquely)
on first principles.

Recently, however, an important breakthrough has been achieved by
the discovery of the so-called inverse kinetic theory (IKT)
approach for
incompressible fluids \cite%
{Ellero2000,Ellero2004,Ellero2005,Tessarotto2006,Tessarotto2007}
which permits a straightforward connection between Eulerian and
Lagrangian descriptions. This is achieved by identifying the
relevant fluid fields,
which are assumed to be defined in a suitable domain $\Omega \subseteq $ $%
\mathbb{R}
^{3}$ (fluid domain), with appropriate moments of a
suitably-defined kinetic distribution function (KDF)
$f(\mathbf{x,}t)$ [with $\mathbf{x=(r,v)}\in \Gamma \mathbf{,}$
$\mathbf{x}$ and $\Gamma $ denoting a suitable state-vector and an
appropriate phase-space] which is assumed to advance in time by
means of Vlasov-type kinetic equation. In such a case, the
time-evolution of the KDF is determined by a kinetic equation
which, written in the Eulerian form, reads
\begin{equation}
Lf(\mathbf{x},t)=0.  \label{Eq.1}
\end{equation}%
Here $f(\mathbf{x},t)$ denotes the Eulerian representation of the
KDF, $L$
is the streaming operator $Lf\equiv \frac{\partial }{\partial t}f+\frac{%
\partial }{\partial \mathbf{x}}\cdot \left\{ \mathbf{X}(\mathbf{x}%
,t)f\right\} ,$ $\mathbf{X}(\mathbf{x},t)\equiv \left\{ \mathbf{v,F}(\mathbf{%
x},t)\right\} $ a suitably smooth vector field, while $\mathbf{v}$ and%
\textbf{\ }$\mathbf{F}(\mathbf{x},t)$ denote respectively
appropriate velocity and acceleration fields. As a main
consequence the approach can in principle be used to determine in
a rigorous way the Lagrangian formulation for arbitrary complex
fluids. Although the choice of the phase space $\Gamma $ is in
principle arbitrary, in the case of incompressible isothermal
fluids, it is found \cite{Ellero2004} that the phase-space $\Gamma
$ can
always be reduced to the direct-product space $\Gamma =\Omega \times V$ (%
\emph{restricted phase-space})$,$ where $\Omega ,V\subseteq $ $%
\mathbb{R}
^{3},$ $\Omega $ is an open set denoted as configuration space of
the fluid (fluid domain) and $V$ is the velocity space. \ This
type of approach (based on a restricted phase-space IKT
formulation) will be adopted also in the sequel.

The main motivation [of this work] is that some of the general
understanding recently achieved in simple flows by means of the
IKT approach could also give a significant contribution to a wider
range of problems. In the sequel, we will concentrate on the issue
of a consistent formulation for fluid dynamics based on a
\emph{phase-space (IKT) description of incompressible fluids,
}whereby its pressure, velocity (and possibly also thermal)
fluctuations are consistently taken into account. In particular we
intend to show that the theory admits a well-defined Lagrangian
formulation. In particular, the corresponding Lagrangian
trajectories can be interpreted in terms of the
\textit{phase-space dynamics} of suitable, \textit{classical
molecules}, i.e., point particles whose dynamics is determined by
the phase-space Lagrangian characteristics. The motion of these
particles (rather than that of the fluid elements), as they are
pushed along erratic trajectories by fluctuations of the
fluid-field gradients (in particular, characterizing the fluid
pressure and temperature), is fundamental to transport and mixing
processes in fluids. In this regard, it is well known that the
interaction between (deterministic and/or turbulent) fluctuations
of the fluid fields and fluid particles still escapes a consistent
theoretical description. Being a subject of major importance for
many environmental, geophysical and industrial applications, the
issue deserves a careful investigation. A key aspect of fluid
dynamics is, therefore, the correct definition of the
(phase-space) Lagrangian dynamics which characterizes
incompressible fluids. The customary approach to the Lagrangian
formulation is based typically on a configuration-space
description, i.e., on the introduction of the configuration-space
Lagrangian
characteristics $\mathbf{r}(t),$ spanning the fluid domain $\Omega .$ Here $%
\mathbf{r}(t)$ denotes the solution of the initial-value problem:%
\begin{equation}
\left\{
\begin{array}{c}
\frac{d\mathbf{r}}{dt}=\mathbf{V}(\mathbf{r},t), \\
\mathbf{r}(t_{o})=\mathbf{r}_{o},%
\end{array}%
\right.  \label{Eq.3}
\end{equation}%
with $\mathbf{r}_{o}$ an arbitrary vector belonging to the closure $%
\overline{\Omega }$ of $\Omega $ and $\mathbf{V}(\mathbf{r},t)$
being the velocity fluid field, to be assumed continuous in
$\overline{\Omega }$ and suitably smooth in $\Omega $. The purpose
of this paper is to achieve a \emph{phase-space Lagrangian
formulation for incompressible thermofluids,} extending the
formulation previously developed, adopting, instead, an
extended-phase-space formulation \cite{Cremaschini2008}. The
present formulation permits to advance in time the relevant fluid
fields by means of
the set of Lagrangian equations defined by the vector field $\mathbf{X}(%
\mathbf{x},t),$ namely
\begin{equation}
\left\{
\begin{array}{c}
\frac{d\mathbf{x}}{dt}=\mathbf{X}(\mathbf{x},t), \\
\mathbf{x}(t_{o})=\mathbf{x}_{o},%
\end{array}%
\right.  \label{Eq.4}
\end{equation}%
where $\mathbf{x}_{o}$ is an arbitrary initial state of
$\overline{\Gamma }$ (closure of the phase-space $\Gamma )$. The
goal of this paper is to show, in particular, that the exact
(phase-space) Lagrangian dynamics can be established based on the
inverse kinetic theory (IKT) for incompressible fluids \emph{in
such a way that the phase space }$\Gamma $\emph{\ coincides
with the direct product space }$\Gamma =\Omega \times V,$\emph{\ }$\Omega $%
\emph{\ being the fluid domain} \emph{and }$V$\emph{\ (velocity
space) the
set }$%
\mathbb{R}
^{3}$. The result appears relevant in particular for the following
reasons: 1) the Lagrangian dynamics here determined permits to
advance in time self-consistently the fluid fields, i.e., in such
a way that they satisfy identically the requires set of fluid
equations. For isothermal fluids, this conclusion is consistent
with the results indicated elsewhere by Tessarotto \textit{et
al.}\cite{Ellero2005}; 2) the Lagrangian dynamics takes into
account the specific form of the phase-space distribution function
which advances in time the fluid fields; 3) the theory permits an
exact description of the motion of those particles immersed in the
fluid which follow the Lagrangian dynamics (classical molecules).
In detail the plan of the paper is as follows. First, in Section
2, an IKT for incompressible thermofluids, adopting a restricted
phase-space description (analogous to that developed in Ref.
\cite{Ellero2005}), is presented. This permits us to
determine the appropriate form of the vector field $\mathbf{X}(\mathbf{x}%
,t). $ For the sake of illustration, we shall consider in
particular the case in which the KDF is identified with a local
Maxwellian distribution function. Second, in Sec.3,\ the new set
of Lagrangian equations is presented, which are proven to advance
uniquely in time the relevant fluid fields of an incompressible
thermofluid.

\section{Eulerian formulation of IKT}

A first key issue is to prove that the IKT approach developed for
isothermal
incompressible fluids \cite%
{Ellero2000,Ellero2004,Ellero2005,Tessarotto2006,Tessarotto2007}
can also be achieved by adopting the same type of \emph{restricted
phase-space formulation,} obtained identifying the phase space
with $\Gamma =\Omega
\times V,$ where $\Omega $ coincides with the fluid domain and $V=$ $%
\mathbb{R}
^{3}$ is a 3D velocity space. For definiteness, here we shall
assume {that
the relevant fluids for a thermofluid, namely }$\left\{ \rho =\rho _{o}>0,%
\mathbf{V},p\geq 0,T>0,S_{T}\right\} ,${\ i.e., respectively the
constant mass density, the fluid velocity, pressure, temperature
and entropy, in the open set $\Omega $ satisfy the so-called
non-isentropic and incompressible Navier-Stokes-Fourier equations
(INSFE), i.e.,}
\begin{eqnarray}
&&\left. \nabla \cdot \mathbf{V}=0,\right.   \label{1} \\
&&\left. \frac{D}{Dt}\mathbf{V}=-\frac{1}{\rho _{o}}\left[ \nabla p-\mathbf{f%
}\right] +\nu \nabla ^{2}\mathbf{V},\right.   \label{2} \\
&&\left. \frac{D}{Dt}T=\chi \nabla ^{2}T+\frac{\nu }{2c_{p}}\left( \frac{%
\partial V_{i}}{\partial x_{k}}+\frac{\partial V_{k}}{\partial x_{i}}\right)
^{2}+\frac{1}{\rho _{o}c_{p}}J\equiv K,\right.   \label{3} \\
&&\left. \frac{\partial }{\partial t}S_{T}\geq 0,\right.
\label{4}
\end{eqnarray}%
where $\frac{D}{Dt}\mathbf{V}$ is the fluid acceleration, $\frac{D}{Dt}=%
\frac{\partial }{\partial t}+\mathbf{V}\boldsymbol{\cdot \nabla }$
the convective derivative and $S_{T}$ a functional, to be suitably
defined, of an appropriate set of fluid fields$.${\ These
equations are assumed to satisfy a suitable initial-boundary value
problem\ (INSFE problem) so that a smooth (strong) solution exists
for the fluid fields }$\left\{ \rho =\rho _{o}>o,\mathbf{V},p\geq
0,T>0\right\} .${\ }Here{\ the notation is standard. Thus,
Eqs.(\ref{1}), (\ref{2}) and (\ref{3}) denote respectively the
so-called \emph{isochoricity condition,} the \emph{forced} \emph{%
Navier-Stokes equation,} written in the Boussinesq approximation,
and the \emph{Fourier equation}.\ As a consequence, in such a case
the force density }$\mathbf{f}$ {reads }$\mathbf{f}=\rho
_{o}\mathbf{g}\left( 1-k_{\rho }T\right) +\mathbf{f}_{1},$ where
the first term represents the
(temperature-dependent) gravitational force density, while the second\ one ($%
\mathbf{f}_{1}$) the action of a possible non-gravitational
externally-produced force. Hence $\mathbf{f}$ can be written also as $%
\mathbf{f}=-\nabla \phi +\mathbf{f}_{R},$ where $\phi =\rho _{o}gz$ and $%
\mathbf{f}_{R}=-\rho _{o}\mathbf{g}k_{\rho }T+\mathbf{f}_{1}$
denote\ respectively the gravitational potential (hydrostatic
pressure) and the non-potential force density. Moreover, in
Eq.(\ref{3}) $J$ is the quantity of heat generated by external
sources per unit volume and unit time. Finally, Eq.(\ref{4})
defines the so-called \emph{2nd} \emph{principle} for the
thermodynamic entropy $S_{T}.$ For its validity in the sequel we
shall
assume that there results everywhere in $\overline{\Omega }\times \overline{{%
I}}$%
\begin{equation}
\int_{\Omega }d\mathbf{r}\left( \chi \nabla ^{2}T+\frac{1}{\rho _{o}c_{p}}%
J\right) \geq 0  \label{fluid with external heating}
\end{equation}%
(\emph{externally heated thermofluid}). In these equations $\mathbf{g,} %
k_{\rho }, \nu ${$, $}$\chi $ and $c_{p}$ are all real constants
which denote respectively the local acceleration of gravity, the
density thermal-dilatation coefficient, the kinematic viscosity,
the thermometric conductivity and the specific heat at constant
pressure. Thus, by taking the divergence of the N-S equation
(\ref{2}), there it follows the Poisson equation for the fluid
pressure $p$ which reads
\begin{equation}
\nabla ^{2}p=-\rho _{o}\nabla \cdot \left( \mathbf{V}\cdot \nabla \mathbf{V}%
\right) +\nabla \cdot \mathbf{f},  \label{5}
\end{equation}%
with $p$ to be assumed non negative and bounded in
$\overline{\Omega }\times \overline{{I}}$. \ In this Section we
shall assume that $f(\mathbf{x,}t)$ is a solution of the Eulerian
kinetic equation (\ref{Eq.1}) defined in a suitable extended
phase-space $\Gamma \times I,$ where $I\subseteq
\mathbb{R}
$ is a suitable time interval. In such a case, we intend to show that $f(%
\mathbf{x,}t)$ can be defined in such a way that the fluid fields $\mathbf{V}%
,p_{1}$ can be identified with its velocity moments $\int\limits_{%
\mathbb{R}
^{3}}d\mathbf{v}G(\mathbf{x,}t)f(\mathbf{x,}t),$ where respectively $G(%
\mathbf{x,}t)=1,\mathbf{v},u^{2}/3$ and $p_{1}$ is the kinetic
pressure to
be defined as:%
\begin{equation}
p_{1}=p_{0}(t)+p-\phi +\rho _{o}T/m.  \label{Eq.6}
\end{equation}%
Here $p_{0}(t)$ (to be denotes as \emph{pseudo-pressure}$\mathbb{\ }$\cite%
{Ellero2005}) is an arbitrary strictly positive and suitably
smooth function defined in $I$. Moreover, $m>0$ is a constant
mass, \textit{whose value remains in principle arbitrary}. In
particular it can be identified with the average mass of the
molecules forming the fluid. Finally, the thermodynamic entropy
$S_{T}$ can be identified with the Shannon statistical entropy
functional $S(f(\mathbf{x,}t))=-\int\limits_{\Gamma }d\mathbf{x}f(\mathbf{x,}%
t)\ln f(\mathbf{x,}t),$ provided the function $p_{0}(t)$ is a
suitably prescribed function and $f(\mathbf{x,}t)$ is strictly
positive in the whole set $\Gamma \times I.$ To reach the proof,
let us first show that, by suitable definition of the vector field
$\mathbf{F}(\mathbf{x},t),$ a particular solution of the IKE
(\ref{Eq.1}) is delivered by the Maxwellian distribution function:
\begin{equation}
f_{M}(\mathbf{x},t)=\frac{\rho _{o}}{\pi ^{2}v_{th,p}^{3}}\exp \left\{ -%
\frac{u^{2}}{v_{th,p}^{2}}\right\} .  \label{Maxwellian}
\end{equation}%
Here $\mathbf{u}=\mathbf{v}-\mathbf{V}(\mathbf{r},t)$ and $v_{th,p}=\sqrt{%
2p_{1}(\mathbf{r},t)/\rho }$ are respectively the relative and the
thermal velocities. The following theorem can immediately be
proven:

\textbf{Theorem - Restricted phase-space INSFE--IKT } \emph{Let us
assume
that:} \emph{1) the INSFE problem admits a smooth strong solution in }$%
\overline{\Gamma }\times I,$ \emph{such that the inequality
(\ref{fluid with
external heating}) is fulfilled}$;$ \emph{2) the vector field }$\mathbf{F}$%
\emph{\ is defined as }%
\begin{equation}
\mathbf{F}(\mathbf{x},t;f)=\mathbf{F}_{0}+\mathbf{F}_{1},
\label{F-1}
\end{equation}%
\emph{where }$\mathbf{F}_{0}\mathbf{,F}_{1}$\emph{\ read respectively}%
\begin{equation}
\mathbf{F}_{0}\mathbf{(x,}t;f)=\frac{1}{\rho _{o}}\left[
\mathbf{\nabla
\cdot }\underline{\underline{{\Pi }}}-\mathbf{\nabla }p_{1}+\mathbf{f}_{R}%
\right] +\mathbf{u}\cdot \nabla \mathbf{V+}\nu \nabla
^{2}\mathbf{V,} \label{F-2}
\end{equation}%
\begin{equation}
\mathbf{F}_{1}\mathbf{(x,}t;f)=\frac{1}{2}\mathbf{u}\left\{ \frac{1}{p_{1}}A%
\mathbf{+}\frac{1}{p_{1}}\mathbf{\nabla \cdot
Q}-\frac{1}{p_{1}^{2}}\left[
\mathbf{\nabla \cdot }\underline{\underline{\Pi }}\right] \mathbf{\cdot Q}%
\right\} +\frac{v_{th}^{2}}{2p_{1}}\mathbf{\nabla \cdot }\underline{%
\underline{\Pi }}\left\{
\frac{u^{2}}{v_{th}^{2}}-\frac{3}{2}\right\} , \label{F-3}
\end{equation}%
\emph{where }%
\begin{equation}
A\equiv \frac{\partial }{\partial t}\left( p_{0}+p\right) -\mathbf{V\cdot }%
\left[ \frac{D}{Dt}\mathbf{V-}\frac{1}{\rho
_{o}}\mathbf{f}_{R}\mathbf{-}\nu \nabla ^{2}\mathbf{V}\right]
+\frac{\rho _{o}K}{m}\equiv \frac{D}{Dt}p_{1}. \label{F-4}
\end{equation}%
\emph{3) in }$\overline{\Omega }\times I$ \emph{the KDF
}$f(\mathbf{x,}t)$
\emph{admits the velocity moments }$G(\mathbf{x,}t)=1,\mathbf{v},u^{2}/3,%
\mathbf{u}\frac{u^{2}}{3}$ \emph{and} $\mathbf{uu;}$ \emph{thus,
we denote in particular} $\mathbf{Q}=\int
d^{3}v\mathbf{u}\frac{u^{2}}{3}f$ \emph{and}
$\underline{\underline{{\Pi }}}=\int d^{3}v\mathbf{uu}f\boldsymbol{;}$ \emph{%
4) the KDF }$f(\mathbf{x,}t)$ \emph{satisfies identically in in }$\overline{%
\Omega }\times I$\emph{\ the constraint equation:}%
\begin{equation}
\rho \equiv \int\limits_{%
\mathbb{R}
^{3}}d\mathbf{v}f(\mathbf{x,}t)=\rho _{o}>0,  \label{constraint on
ro_0}
\end{equation}%
\emph{where }$\rho _{o}$\emph{\ is a positive constant;} \emph{5)
the
entropy integral }$S(f(\mathbf{x,}t_{0}))=-\int\limits_{\Gamma }d\mathbf{x}f(%
\mathbf{x,}t_{0})\ln f(\mathbf{x,}t_{o})$ \emph{exists;} \emph{6)
in the time interval }$I$\emph{\ the pseudo-pressure
}$p_{0}(t)$\emph{\ is defined
so that there results identically:}%
\begin{equation}
\int_{\Omega }d\mathbf{r}\frac{1}{p_{1}}\left[ \frac{\partial }{\partial t}%
p_{1}+\mathbf{\nabla \cdot Q-}\frac{1}{p}\mathbf{\nabla }p\mathbf{\cdot Q}%
\right] =0.  \label{constarint on p0}
\end{equation}%
\emph{It follows that :} \emph{A) the local Maxwellian
distribution function (\ref{Maxwellian}) is a particular solution
of the IKE (\ref{Eq.1}) if an
only if the fluid fields }$\left\{ \rho =\rho _{o}>0,\mathbf{V},p,T\right\} $%
\emph{\ satisfy the fluid equations (\ref{1})-(\ref{3}). In such a
case
there results identically }$\mathbf{Q=0,}$\textbf{\ }$\underline{\underline{{%
\Pi }}}=\underline{\underline{0}}$\emph{;} \emph{B) for an arbitrary KDF }$f(%
\mathbf{x},t),$ \emph{which satisfies identically assumptions
4)-6), the velocity-moment equations obtained by taking the
weighted velocity integrals
of Eq.(\ref{Eq.1}) with the weights }$G(\mathbf{x},t)=1,\mathbf{v},u^{2}/3$%
\emph{\ deliver identically the same fluid equations
(\ref{1})-(\ref{3});} \emph{C) the Shannon entropy }$S\left(
f\right) $\emph{\ is a monotonic function of time, i.e., there
results (}$\forall t\in I$\emph{)}$:$\emph{\ \ }
\begin{equation}
\frac{\partial }{\partial t}S\left( f\right) \geq 0
\end{equation}%
\emph{(H-theorem), where there results in particular for
isothermal fluids (see also \cite{Cremaschini2008})}
$\frac{\partial }{\partial t}S\left(
f\right) =0$ \emph{(constant H-theorem). This permits us to identify } $%
S_{T}=S\left( f\right) $\emph{\ so that the 2nd principle [i.e.,
the inequality (\ref{4})] is satisfied too.}\newline PROOF First,
let us assume that a strong solution of the INSFE problem
exists which satisfies identically Eqs.(\ref{1})-(\ref{3}) in the set $%
\Omega \times I$. In such a case it is immediate to prove that $f_{M}(%
\mathbf{x},t)$ is a particular solution of the inverse kinetic equation (\ref%
{Eq.1})$.$ The proof follows upon invoking
Eqs.(\ref{F-1})-(\ref{F-4}) for the vector field
$\mathbf{F}(\mathbf{x},t;f)$ and by direct substitution of the
distribution $f_{M}(\mathbf{x},t)$ in the same equation
(Proposition A). Instead, if we assume that in $\Gamma \times I,$
$f\equiv f_{M}(\mathbf{x},t) $ is a particular solution of the
inverse kinetic equation, which fulfills identically the
constraint equation (\ref{constraint on ro_0}), it follows that
the fluid fields $\left\{ \rho =\rho _{o}>0,\mathbf{V},p,T\right\}
$ are necessarily solutions of the INSFE equations. This can be
proven either:
a) by direct substitution of $f\equiv f_{M}(\mathbf{x},t)$ in Eq.(\ref{Eq.1}%
) (Proposition A); b) by direct evaluation of the velocity moments
of the same equation for $G(\mathbf{x},t)=1,\mathbf{v},u^{2}/3$
(Proposition B). In fact, thanks to assumptions 1-6 the first two
moment equation coincide
respectively with the isochoricity and Navier-Stokes equations [Eqs. (\ref{1}%
) and (\ref{2})]. As a consequence the energy equation is also
satisfied by
the fluid fields $\mathbf{V}$ and $p,$ namely there results identically in $%
\Omega \times I$%
\begin{equation}
\mathbf{V\cdot }\left[ \frac{D}{Dt}\mathbf{V}+\frac{1}{\rho
_{o}}\left[
\nabla \left( p-\phi \right) -\mathbf{f}_{R}\right] -\nu \nabla ^{2}\mathbf{V%
}\right] =0.
\end{equation}%
Therefore, the third moment equation delivers the Fourier equation [Eq.(\ref%
{3})]. The same proof (for Proposition B) is straightforward if $f\neq f_{M}(%
\mathbf{x},t).$ This is reached again imposing the same constraint equation (%
\ref{constraint on ro_0}) on first velocity-moment of the
distribution function $f.$ Finally, the proof of Proposition C is
achieved by imposing on $p_{0}(t)$ the constraint equation
(\ref{constarint on p0}) and invoking assumptions 1) and 5). It
follows that the thermodynamic entropy can be
identified with the Shannon entropy, i.e., letting for all $t\in I,$ $%
S_{T}(t)=S\left( f_{M}(\mathbf{x},t)\right) .$ As a final remark,
we point out that by properly prescribing the initial and boundary
conditions for the KDF, one can show that also the appropriate
initial and boundary condition
for the fluid fields can be satisfied (see related discussion in Ref. \cite%
{Ellero2005}). Furthermore, by construction, due to the constraint (\ref%
{constraint on ro_0}),
$\widehat{f}(\mathbf{x},t)=f(\mathbf{x},t)/\rho _{o}$
is a velocity-space probability density, i.e., there results identically in $%
\overline{\Omega }\times I,\int\limits_{%
\mathbb{R}
^{3}}d\mathbf{v}\widehat{f}(\mathbf{x,}t)=1$. \

\section{Lagrangian formulation of IKT}

The results of the previous Section permit us to formulate in a
straightforward way also the equivalent Lagrangian form of the
inverse kinetic equation \cite{Ellero2005}. The Lagrangian
formulation is achieved in two steps: a) by identifying a suitable
dynamical system (here denoted as \emph{INSFE dynamical system}),
which determines uniquely the time-evolution of the kinetic
probability density prescribed by IKT. Its flow defines a
family of phase-space trajectories, here denoted as phase-space \emph{%
Lagrangian paths} (LP's); b) by proper parametrization in terms of
the LP's the KDF and the inverse kinetic equation, the explicit
solution of the initial-value problem defined by the inverse
kinetic equation (\ref{Eq.1}) is determined. First, we notice that
- in view of the THM - it is obvious that the LP's must be
identified with the phase-space trajectories of a
classical dynamical system $\mathbf{x}_{o}\rightarrow \mathbf{x}%
(t)=T_{t,t_{o}}\mathbf{x}_{o}$ generated by the vector field $\mathbf{X}(%
\mathbf{x},t).$ Hence we shall assume that the initial-value problem (\ref%
{Eq.4}), which is realized by the equations%
\begin{equation}
\left\{
\begin{array}{c}
\frac{d}{dt}\mathbf{r}(t)=\mathbf{v}(t), \\
\frac{d}{dt}\mathbf{v}(t)=\mathbf{F}(\mathbf{r}(t),t;f), \\
\mathbf{r}(t_{o})=\mathbf{r}_{o}, \\
\mathbf{v}(t_{o})=\mathbf{v}_{o},%
\end{array}%
\right.  \label{Eq.4'}
\end{equation}%
defines a suitably smooth diffeomeorphism. Here, by construction:

\begin{itemize}
\item denoting by $\mathbf{x}(t)=\chi (\mathbf{x}_{o},t_{o},t)$ the solution
of the initial-value problem (\ref{Eq.4'}), $\mathbf{x}_{o}=\chi (\mathbf{x}%
(t),t,t_{o})$ is its inverse. Both are assumed to be suitably
smooth functions of the relevant parameters;

\item both $\mathbf{x}(t)=\chi (\mathbf{x}_{o},t_{o},t)$ and $\mathbf{x}%
_{o}=\chi (\mathbf{x}(t),t,t_{o})$ \ identify admissible LP's of
the dynamical;

\item $\mathbf{r}(t)$ is the Lagrangian trajectory which belongs to the
fluid domain $\Omega $;

\item $\mathbf{v}(t)$ and\ $\mathbf{F}(\mathbf{r}(t),t;f)$ are respectively
the Lagrangian velocity and acceleration, both spanning the vector space $%
\mathbb{R}
^{3}.$ In particular, \ $\mathbf{F}(\mathbf{r}(t),t;f),$ which is
defined by Eqs.(\ref{F-1})-(\ref{F-4}), and depends functionally
on the kinetic probability density $f(\mathbf{x},t),$ is the
\emph{Lagrangian acceleration
which corresponds to an arbitrary kinetic probability density} $f(\mathbf{x}%
,t);$

\item $f(\mathbf{x},t)$ is a particular solution of the inverse kinetic
equation \ref{Eq.1} which is subject to the assumptions imposed by
the THM.
\end{itemize}

It follows that the Jacobian [$J(\mathbf{x}(t),t)=\left\vert
\frac{\partial
\mathbf{x}(t)}{\partial \mathbf{x}_{o}}\right\vert $] of the map $\mathbf{x}%
_{o}\rightarrow \mathbf{x}(t),$ which is generated by
Eq.(\ref{Eq.4'}) for a
generic distribution function $f(\mathbf{x},t)$ of this type, reads%
\begin{equation}
J(\mathbf{x}(t),t)=\left\vert \frac{\partial \mathbf{x}(t)}{\partial \mathbf{%
x}_{o}}\right\vert =\exp \left\{ \int\limits_{t_{o}}^{t}dt^{\prime }H(%
\mathbf{x(}t^{\prime }),t^{\prime })\right\} ,
\end{equation}

where $H(\mathbf{x},t)\equiv \frac{3}{2}\left[ \frac{1}{p_{1}}A\mathbf{+}%
\frac{1}{p_{1}}\mathbf{\nabla \cdot Q}-\frac{1}{p_{1}^{2}}\left[ \mathbf{%
\nabla \cdot }\underline{\underline{\Pi }}\right] \mathbf{\cdot Q}\right] +%
\frac{1}{p_{1}}\mathbf{u\cdot \nabla \cdot
}\underline{\underline{\Pi }}.$
Instead, in the case in which there results identically $f\equiv f_{M}(%
\mathbf{x},t),$ the Jacobian reduces to
\begin{equation}
J(\mathbf{x}(t),t)=\left\vert \frac{\partial \mathbf{x}(t)}{\partial \mathbf{%
x}_{o}}\right\vert =\frac{v_{th,p}^{3}(t)}{v_{th,p}^{3}(t_{o})}\exp \left\{ -%
\frac{u^{2}(t_{o})}{v_{th,p}^{2}(t_{o})}+\frac{u^{2}(t)}{v_{th,p}^{2}(t)}%
\right\} .
\end{equation}

Second, it is immediate to prove that the kinetic equation in the
Lagrangian representation can be written in the form

\begin{equation}
J(\mathbf{x}(t),t)f(\mathbf{x}(t),t)=f(\mathbf{x}_{o},t_{o})\equiv f_{o}(%
\mathbf{x}_{o})  \label{Eq.2}
\end{equation}%
where $f(\mathbf{x}(t),t)$ is the Eulerian representation of the KDF and $%
f_{o}(\mathbf{x}_{o})$ is{\ a suitably smooth initial KDF}.
Eq.(\ref{Eq.2}) manifestly implies also the time evolution of
$f(\mathbf{x}(t),t)$ in terms of the initial distribution
function:
\begin{equation}
f(\mathbf{x}(t),t)=\frac{1}{J(\mathbf{x}(t),t)}f_{o}(\chi (\mathbf{x}%
(t),t,t_{o})).  \label{Eq.7}
\end{equation}%
From the \textit{mathematical standpoint} main consequences of the
theory are that: 1) the Lagrangian formulation (of IKT) is
uniquely specified by
the proper definition of a suitable family of phase-space LP's; 2) Eq.(\ref%
{Eq.7}) uniquely specifies the time-evolution of the Eulerian KDF, $f(\mathbf{%
x}(t),t),$ which is represented in terms of the initial
distribution function $f_{o}(\mathbf{x}_{o})$ and the LP's defined
by the INSFE dynamical
system; 3) the time-evolution of the fluid fields $\left\{ \rho =\rho _{o}>0,%
\mathbf{V},p\geq 0,T>0,S_{T}\right\} $ is uniquely specified via the KDF $f(%
\mathbf{x}(t),t);$ 4) Eq.(\ref{Eq.7}) also provides the connection
between
Lagrangian and Eulerian viewpoints. In fact the Eulerian KDF, $f(\mathbf{x}%
,t),$ is simply obtained from Eq.(\ref{Eq.7}) by letting $\mathbf{x}=\mathbf{%
x}(t)$ in the same equation. As a result, the Eulerian and
Lagrangian formulations of IKT, and hence of the underlying moment
(i.e., fluid)
equations, are manifestly equivalent$.$ From the \textit{physical viewpoint}%
, it is worth mentioning the LP's here defined can be interpreted
as phase-space trajectories of the particles of the fluid, to be
considered as a set of "classical molecules", i.e., point
particles with prescribed mass, which interact only via the action
of a suitable mean-field force kind. The ensemble motion of these
particles has been defined in such a way that it uniquely
determines the time evolution \textit{both} of the kinetic
distributions functions \textit{and} of the relevant fluid fields
which characterize the thermofluid.

\section{Concluding remarks}

In this Note an inverse kinetic theory has been developed for the
INSFE problem based on a restricted phase-space representation, \
i.e., in which the phase space of the kinetic description is
identified with the direct product \emph{\ }$\Gamma =\Omega \times
V$ defined in such a way that\emph{\ }$\Omega $\emph{\ }coincides
with the fluid. We have shown that equivalent IKT approaches can
be formulated both in the Eulerian and Lagrangian viewpoints. The
result appears relevant for several reasons, in particular: 1) the
Lagrangian dynamics here determined permits to advance in time
self-consistently the fluid fields, i.e., in such a way that they
satisfy identically the requires set of fluid equations. For
isothermal fluids, this conclusion is consistent with the results
indicated elsewhere by Tessarotto \textit{et
al.}\cite{Ellero2005}; 2) the Lagrangian dynamics, defined in the
configuration space of the fluid (fluid domain), takes into
account the specific form of the phase-space distribution function
which advances in time the same fluid fields; 3) the theory
permits an exact description of the motion of the "classical
molecules" which follow the Lagrangian dynamics. Finally, in our
view the formulation here presented is promising in turbulence
theory. In fact, the connection between Eulerian and Lagrangian
phase-space descriptions based on a IKT approach, here established
on rigorous grounds, represents a potentially useful new
development.

\section*{Acknowledgments}
Work developed in cooperation with the CMFD Team, Consortium for
Magneto-fluid-dynamics (Trieste University, Trieste, Italy). \
Research developed in the framework of the MIUR (Italian Ministry
of University and Research) PRIN Programme: \textit{Modelli della
teoria cinetica matematica nello studio dei sistemi complessi
nelle scienze applicate}. The support COST Action P17 (EPM,
\textit{Electromagnetic Processing of Materials}) and GNFM
(National Group of Mathematical Physics) of INDAM (Italian
National Institute for Advanced Mathematics) is acknowledged.

\section*{Notice}
$^{\S }$ contributed paper at RGD26 (Kyoto, Japan, July 2008).
\newpage




\begin{thebibliography}{BIBTEX}
\bibitem{Richardson} L. F. Richardson, Proc. R. Soc. London A 110, 709
(1926).

\bibitem{Pope2000} S.B. Pope, \textit{Turbulent flows}, Cambridge University
Press, p.463 (2000).

\bibitem{Ellero2000} M. Ellero and M. Tessarotto, Bull. Am Phys. Soc.
\textbf{45 }(9), 40 (2000).

\bibitem{Ellero2004} M. Tessarotto and M. Ellero, RGD24 (Italy, July 10-16,
2004), AIP Conf. Proc. \textbf{762}, 108 (2005).

\bibitem{Ellero2005} M. Ellero and M. Tessarotto, Physica A \textbf{355},
233 (2005).

\bibitem{Tessarotto2006} M. Tessarotto and M. Ellero, Physica A \textbf{373}%
, 142 (2007); arXiv: physics/0602140.

\bibitem{Tessarotto2007} M. Tessarotto and M. Ellero, Proc. 25th RGD
(International Symposium on Rarefied gas Dynamics, St. Petersburg,
Russia, July 21-28, 2006), Ed. M.S. Ivanov and A.K. Rebrov
(Novosibirsk Publ. House of the Siberian Branch of the Russian
Academy df Sciences), p.1001; arXiv:physics/0611113 (2007).

\bibitem{Cremaschini2008} C. Cremaschini and M.Tessarotto, \textit{Inverse
kinetic theory for incompressible thermofluids}, contributed paper
at RGD26 (Kyoto, Japan, July 2008); arXiv:0806.4546 (2008).


\end{thebibliography}
\end{document}